\documentclass[useAMS,usenatbib]{mn2e}
\usepackage{epsfig}
\usepackage{graphicx}

\begin{document}

\title[Spin evolution of neutron stars with alignment or
counteralignment]
{Be born slow or die fast: spin evolution of neutron stars with alignment or
counteralignment}

\author[S.A. Eliseeva, S.B. Popov, V.S. Beskin]{S. A. Eliseeva$^1$, S.B. Popov$^2$ and V.S.
Beskin$^1$
\thanks{E-mail: sa\_eliseeva@comcast.net (SAE); polar@sai.msu.ru (SBP); beskin@lpi.ru (VSB)}\\
$^1$ Lebedev Physical Institute, Russian Academy of Science, Leninski pr. 53, Moscow 119991, Russia
\\
$^2$ Sternberg Astronomical Institute, Universitetski pr. 13, Moscow 119992,
Russia}
\date{Accepted ...... Received ......; in original form......}

\pagerange{\pageref{firstpage}--\pageref{lastpage}} \pubyear{2006}

\maketitle

\label{firstpage}

\begin{abstract}
We revisit the case of magneto-rotational evolution of neutron stars
(NSs) with accounting for changes of the angle $\chi$ between spin
and magnetic axes. This element of the evolution of NSs is very
important for age estimates and population modeling, but usually it
is neglected. In the framework of two models of energy losses we
demonstrate that unless NSs are born with the inclination angle
$\chi_0$ very close (within $\sim 1^\circ$) to the position of
maximal losses, pulsars with short initial periods quickly reach
regions of small energy losses (aligned rotators in the case of
magneto-dipole losses and orthogonal rotators -- in the case of
current losses) without significant spin-down. This means that
either most of known NSs should be born with relatively long periods
close to presently observed values, or the initial inclination
angles should be large (close to $90^{\circ}$) for the magnetodipole
model and very small (close to $0^{\circ}$) for the current losses
model, or that both models cannot be applied for the whole
evolutionary track of a typical NS. In particular, magnetar
candidates should be born with periods close to the observed ones
within the scope of both models of energy losses. We discuss how
these considerations can influence population synthesis models and
age estimates for different types of isolated NSs. Focusing on the
model of current losses we illustrate our conclusion with
evolutionary track reconstructions on the $P$~--~$\sin \chi$ plane.
We conclude, that most probably both existing models of energy
losses -- magneto-dipole and longitudinal current -- require serious
modifications, on the other hand, estimates obtained under the
standard assumption of $\sin \chi=1=$~const does not have solid
theoretical ground and should be taken with care.
\end{abstract}
\begin{keywords}
stars: neutron  --- stars: evolution --- pulsars: general
\end{keywords}

\section{Introduction}
\label{intro}

Radio pulsars have been discovered nearly 40 years ago
\citep{bell67}, and very quickly they were recognized as magnetized
rotating neutron stars spinning down and dissipating their
rotational energy. Despite of all the progress in the theory of
magneto-rotational evolution of isolated neutron stars, several
important questions still remain unclear. The uncertainty concerning
the evolution of the angle $\chi$ between the magnetic axis and the
spin axis of an isolated neutron star is one of them, although the
angle $\chi$ significantly affects the rate of energy release for
all mechanisms of radio pulsars emission. At that, among the basic
parameters, which determine astrophysical manifestations of radio
pulsar, this one is the least popular, and its evolution is usually
neglected.

We discuss evolution and estimate ages of isolated neutron stars
taking into account changes of the angle between the magnetic and
the spin axes. We consider the magnetodipole model and the model
of pulsar radio emission based on current losses in the
magnetosphere of a neutron star, focusing on the latter. It should
be emphasized here that the evolution of an inclination angle is a
key parameter for both models. Unfortunately, many authors either
completely disregard the evolution of an inclination angle  or at
best consider its result to be identical to the magnetic field
decay (see, for example, \citealt{reg2001,fk06}, and references
therein).

In the following section of this paper we consider the
magnetodipole and current losses models of energy release of radio
pulsars. There the basic information about the models themselves
and their main contradictions with observations are listed. The
expression for the characteristic ages of isolated neutron stars
with the account of evolution of an inclination angle is
presented. As a result, the ages of isolated neutron stars
estimated by the obtained formulas for different initial
parameters (spin periods and inclination angles) are listed in
Tables 1 and 2. In the third section we reconstruct possible
evolutionary tracks of XDIN RX J0720.4-3125 and AXP 1E2259+586
varying parameters and assuming the model based on current losses
is valid. In the forth sections we give a discussion. In
particular, as the account for changes of $\chi$ mainly influence
population synthesis models and age estimates, we discuss these
topics. Finally, we present our conclusions.

\section{The role of the evolution of inclination angle in the age estimates on the active radio pulsar stage}
\label{ages_vs_angles}

%One of the key parameters of an isolated neutron star is its angle
%between the magnetic and spin axes. Unfortunately, from the
%observational point of view there exists only little information
%about the inclination angles immediately after the birth of a
%neutron star \citep{gr2006}(Geppert~\&~Rheinhardt 2006). The problem
%arises from the poor observability of newborn neutron stars, which
%are surrounded by opaque medium. There are a few papers in which the
%initial inclination angles of some radio pulsars are estimated. For
%example, (Harding et al. 2003) find that the initial inclination
%angles of all considered by them fast rotating radio pulsars are
%less than $30^{\circ}$. At this point we want to note that the small
%initial inclination angles are typical for the so-called model based
%on the current losses \citep{bgi1993}.

In this section we estimate duration of active radio pulsar stage
for isolated neutron stars taking into account the evolution of
the angle between the magnetic and the spin axes. We consider two
models of energy release mechanism: the magnetodipole model and
current losses model. Everywhere in the paper we neglect the
possibility that magnetic field decays, and that there are any
non-electromagnetic effects like those described, for example, by
\cite{macy1974, m2001, w2003, r2006}.

 The primary point of our calculations is
that the ratio $f(\chi)/P$, where $P$ -- is spin period, and
$f(\chi)=\cos \chi$ for the magnetodipole model and $f(\chi)=\sin
\chi$ for the current losses model, is invariant during the
evolution of a neutron star on the radio pulsar stage. Therefore,
in order to estimate the characteristic age of an isolated neutron
star in scope of one of the models we do not need to know the
initial inclination angle. The inclination angle and the spin
period of a neutron star at the present time and its initial spin
period are enough to define the initial inclination angle for each
model of energy release. The validity and implications of the
invariants are discussed throughout this paper.

\subsection{Magnetodipole model}
\label{magnetodipole}

Energy losses of an isolated radio pulsar due to  radiation can be
associated with two different scenarios. The first one is based on
magnetic dipole braking, and we discuss it in this subsection. In
this case, assuming a perfect dipole rotating in vacuum, the
energy losses can be estimated by the so-called magnetodipole
formula \citep{p1967}:

\begin{equation}
W_{\mathrm{md}}=-J \Omega \dot{\Omega}
= \frac{1}{6} \frac{B_0^2 \Omega^4 R^6}{c^3}
\sin^2 \chi.
\label{Wmd}
\end{equation}
Here $J \sim 2/5MR^2$ is a moment of inertia of a neutron star,
$\chi$ is the angle between the spin and the magnetic axes, and
$\Omega=2\pi /P$ is the angular velocity of a neutron star. The
magnetic field is defined as $B_0=2\mu/R^3$, i.e. this is the field
on a pole. The magnetodipole losses for ordinary pulsars are
approximately $10^{31}$~--~$10^{34} \mbox{ erg}\,  \mbox{ s}^{-1}$.
A simple estimate of a lifetime of a normal radio pulsar with a spin
period $P=1$~s and period derivative $\dot{P}=10^{-15}$~s~s$^{-1}$
gives us $\tau_{\mathrm{ch}}=P/ 2 \dot{P} \approx 10^7$~yrs. One can
see that this result is only a rough approximation; for more
accurate estimate of the age of a radio pulsar the evolution of the
angle between the magnetic and the spin axes should be taken into
account. According to (\ref{Wmd}), the slow-down rate of a radio
pulsar for the magnetodipole model can be defined as follows:

\begin{equation}
\dot{P}= 0.24 \times 10^{-15}B_{12}^{2}P^{-1}\sin^{2} \chi.
\label{dotPmd}
\end{equation}
 In the scope of the magnetodipole model the radiation-reaction
torque is perpendicular to the magnetodipole moment of a neutron
star. Therefore, the component of angular velocity $\Omega \cos
\chi$ parallel to the magnetodipole moment (and so perpendicular to
the deceleration moment) remains constant \citep*{dg1970,
michel1991}, i.e. the ratio
\begin{equation}
I_{\mathrm{md}}=\frac{\cos \chi}{P}
\label{Invmd}
\end{equation}
is invariant during the evolution of a neutron star at the active
radio pulsar stage (since birth of a neutron star and up to the
moment when it crosses the death line). Thus, the angle between the
spin and the magnetic axes during the evolution tends
to~$0^{\circ}$. \cite{dg1970} emphasize that this model has the
following implication: if radio pulsars are born as fast rotators
and are decelerated only by dipole radiation to spin periods close
to one second then the initial inclination angles must be very close
to $90^{\circ}$.

Here we outline the problem that most of the authors disregard as
they do not take into account the evolution of the angle between
the magnetic axis and the spin axis. It is often accepted,
especially in population synthesis models, that radio pulsars are
born as fast rotators ($P_0 \sim 10-30$~ms). As there is no
information about the initial distribution of radio pulsars in the
inclination angle \citep{tm1998}, most of the authors assume the
distribution in the initial inclination angle to be equiprobable,
or just ignore this question. Eq.~(\ref {Invmd}) defines the
evolutionary track of a newborn radio pulsar with given $P_0$ and
$\chi_0$ till it reaches the death line, which for the
magnetodipole model is generally approximated by the following
simple equation (see, for example, \citealt{bhat1992}):

\begin{equation}
 \frac{B}{P^2}=0.34 \times 10^{12} \, \mathrm{G} \, \mathrm{s}^{-2}.
\label{dlmd}
\end{equation}
It should be noted here that the death line approximation
(\ref{dlmd}) is valid for a neutron star with a perfect dipole
magnetic field and with the radius $R=10^6$ cm, the moment of
inertia $J=10^{45}$~g~cm$^2$, and the inclination angle $\sin \chi
\sim 1$, i.e. $B$ is calculated out of $\dot P$ under these
assumptions. The lack of dependence of the death line on the angle
$\chi$ should be considered as an important disadvantage of the
usual approach in the context of the magnetic dipole braking model.
In the model of current losses such dependence is taken into account
(see below). Accordingly,  the accuracy of age estimates  is often
lower for the magnetodipole model than for the model based on
current losses. As a result of (\ref{dlmd}), for the period
$P_{\mathrm{dl}}$, at which neutron star is leaving the radio pulsar
stage (the death line crossing), we have:

\begin{equation}
P_{\mathrm{dl}}=1.7 \,B_{12}^{0.5} \, \mathrm{s},
\label{Pdlmd}
\end{equation}
here $B_{12}=B/10^{12}$~G.

 Thus, if
\begin{equation}
P_{\mathrm{dl}} > \frac{P_0}{\cos \chi_0},
\label{conditionmd}
\end{equation}
an isolated neutron star becomes an aligned rotator before it can
reach the death line and leave the stage of radio emission. As one
can see, this way of evolution should be typical for most of the
radio pulsars with short initial spin periods, except for those with
the initial angles between the magnetic and the spin axes very close
to $90^{\circ}$. For example, let us consider an evolutionary track
of the neutron star with the initial spin period $P_0=20$~ms,
initial angle between the spin and magnetic axes $\chi_0=45^{\circ}$
and magnetic field $B_0=10^{12}$~G. Eq.~(\ref{Invmd}) allows us to
estimate the period at which the neutron star becomes an aligned
rotator: $P_{\mathrm{al}} \approx 28$~ms.

As a result, for the magnetodipole model we conclude the following: if the
invariant (\ref{Invmd}) is valid, and radio pulsars are born as fast
rotators, and their initial angle distribution is equiprobable, then most of
radio pulsars should become aligned rotators long before they pass the death
line. Thus, we should expect the distribution in observable angle between
the magnetic and the spin axes of active radio pulsars to have a peak close to
$0^{\circ}$. Otherwise, an essential part of radio pulsars should initially
have long spin periods \citep*{bgi1993}.
%The statistics of active radio pulsars is discussed below.

Further, we can estimate a lifetime of an isolated neutron star as
an active radio pulsar:

\begin{equation}
\tau_{\mathrm{md}}[{\mathrm
s}]=\int_{P_0}^{P_{\mathrm{dl}}}\frac{dP} {2.4 \times
10^{-16}B_{12}^{2}P^{-1} \left(1- \cos^2 \chi_0(P^2)/(P_0^2)
\right)}, \label{intmd}
\end{equation}
where $P_0$ and $\chi_0$ are the initial spin period and the angle
between the spin and the magnetic axes of a radio pulsar,
respectively, and $P_{\mathrm{dl}}$ is a period at which a radio
emission turn-off occurs (a pulsar crosses the death line). If an
inclination angle $\chi$ reaches $0^{\circ}$ before a pulsar crosses
the death line, the period of the transition to aligned rotators
$P_{\mathrm{al}}$ should be used instead of $P_{\mathrm{dl}}$.

Substituting Eq.~(\ref{Invmd}) into (\ref{intmd}) and integrating, we
obtain:
\begin{equation}
\tau_{\mathrm{md}}=2 \times 10^{15}B_{12}^{-2}
\left( \frac{P_0}{\cos \chi_0} \right)^{2} \ln \left|
\frac{1 - \cos^{-2} \chi_0}{P_{\mathrm{dl}}^2 P_0^{-2} - \cos^{-2} \chi_0}
\right| \, \mathrm{s}.
\label{taumd}
\end{equation}
Therefore, assuming the initial spin period $P_0=20$~ms, initial
angle between the spin and the magnetic axes $\chi_0=89.6^{\circ}$
and magnetic field $B_0=10^{12}$~G, from Eq.~(\ref{taumd}) we
obtain an estimate of the age of an isolated neutron star at the
moment it crosses the death line as $\tau_{\mathrm{md}}=226$~Myr.
If we assume $\chi_0=45^{\circ}$, the age of the radio pulsar
which enters the stage of aligned rotators can be estimated as
$\tau_{\mathrm{al}}=0.16$~Myr. Table~1 contains the estimates of
duration of the radio pulsar stage for different parameters (such
as magnetic fields, initial spin periods and initial inclination
angles of radio pulsars).

It should be noted here that the age of a neutron star estimated
within the scope of magnetodipole model by (\ref{taumd})
significantly depends on the death line. For example, if the
coefficient in (\ref{Pdlmd}) is twice less, then the lifetime of a
typical Crab-like pulsar at the radio pulsar stage estimated by
(\ref{taumd}) is reduced by at least the factor of 4. At that, the
smaller inclination angles $\chi$ at which the pulsar crosses the
death line gives the larger decrease in the estimated lifetimes
with the coefficient reduction in (\ref{Pdlmd}). As the spin-down
rate of a neutron star decreases near the death line (due to the
factor $\sin ^2 \chi$ in Eq.~\ref{dotPmd}), the longest part of a
lifetime at the radio pulsar stage is the period when a neutron
star stays near the death line. Thus, as we do not take into
account the inclination angle $\chi$ for the death line in the
scope of the magnetodipole model Eq.(\ref{taumd}) only roughly
approximates the age of a neutron star. It is most likely that the
more accurate expression for the death line can reduce the age
estimates by a factor of a few.

\begin{table*}
 \centering
 \begin{minipage}{140mm}
  \caption{Age estimates for the magnetodipole model. $P_0$ and $\chi_0$ are initial parameters of a neutron star,
  $P_{\rm dl}$ and $\tau_{\rm md}$ are the spin period and the estimated age of a neutron star on the death line,
  $P_{\rm al}$ and $\tau_{\rm al}$ are the spin period and the estimated age of a neutron star at the moment of transition to the aligned rotators.}
  \begin{tabular}{@{}lrrllll@{}}
  \hline
B, G & $P_0$, s & $\chi_0$, deg. & $P_{\rm dl}$, s & $\tau_{\rm md}$, Myr & $P_{\rm al}$, s & $\tau_{\rm al}$, Myr \\
\hline
 $10^{12}$ & 0.02 & 30   & -     & -      & 0.023 & 0.1              \\
 $10^{13}$ & 0.02 & 30   & -     & -      & 0.023 & $10^{-3}$ \\
 $10^{14}$ & 0.02 & 30   & -     & -      & 0.023 & $10^{-5}$ \\
 $10^{12}$ & 0.02 & 45   & -     & -      & 0.028 & 0.2             \\
 $10^{13}$ & 0.02 & 45   & -     & -      & 0.028 & $2 \times 10^{-3}$ \\
 $10^{14}$ & 0.02 & 45   & -     & -      & 0.028 & $2 \times 10^{-5}$ \\
 $10^{12}$ & 0.02 & 60   & -     & -      & 0.04 & 3.6             \\
 $10^{13}$ & 0.02 & 60   & -     & -      & 0.04 & 0.04 \\
 $10^{14}$ & 0.02 & 60   & -     & -      & 0.04 & $4 \times 10^{-4}$ \\
 $10^{12}$ & 0.02 & 89.6 & 1.7   & 226    & -     & -                 \\
 $10^{13}$ & 0.02 & 89.7 & 5.4   & 46     & -     & -                 \\
 $10^{14}$ & 0.02 & 89.9 & 17    & 2.7    & -     & -                 \\
 $10^{12}$ & 0.5  & 30   & -     & -      & 0.577 & 112              \\
 $10^{13}$ & 0.5  & 30   & -     & -      & 0.577 & 1.1              \\
 $10^{14}$ & 0.5  & 30   & -     & -      & 0.577 & 0.01             \\
 $10^{12}$ & 0.5  & 45   & -     & -      & 0.707 & 235              \\
 $10^{13}$ & 0.5  & 45   & -     & -      & 0.707 & 2.4              \\
 $10^{14}$ & 0.5  & 45   & -     & -      & 0.707 & 0.02             \\
 $10^{12}$ & 0.5  & 60   & -     & -      & 1     & 2222              \\
 $10^{13}$ & 0.5  & 60   & -     & -      & 1     & 22              \\
 $10^{14}$ & 0.5  & 60   & -     & -      & 1     & 2.2             \\
 $10^{12}$ & 0.5  & 89.6 & 1.7   & 167    & -     & -                 \\
 $10^{13}$ & 0.5  & 89.7 & 5.4   & 18     & -     & -                 \\
 $10^{14}$ & 0.5  & 89.9 & 17    & 1.8    & -     & -                 \\
\hline
\end{tabular}
\end{minipage}
\end{table*}

\subsection{Longitudinal current losses model}
\label{currents}

\begin{table*}
 \centering
 \begin{minipage}{140mm}
  \caption{Age estimates for the model based on longitudinal current losses.
  $P_0$ and $\chi_0$ are initial parameters of a neutron star,
  $P_{\rm dl}$ and $\tau_{\rm cl}$ are the spin period and the
  estimated age of a neutron star on the death line, $P_{\rm ort}$
  and $\tau_{\rm ort}$ are the spin period and the estimated age of a neutron star
  at the moment of crossing the boundary of the region of orthogonal rotators.}
  \begin{tabular}{@{}lrrllll@{}}
  \hline
B, G & $P_0$, s & $\chi_0$, deg. & $P_{\rm dl}$, s & $\tau_{\rm cl}$, Myr & $P_{\rm ort}$, s & $\tau_{\rm ort}$, Myr \\
\hline
 $10^{12}$ & 0.02 & 0.4& 0.98  & 34   & -     & -   \\
 $10^{13}$ & 0.02 & 0.3& 2.95  & 4.2  & -     & -   \\
 $10^{14}$ & 0.02 & 0.1& 9.51  & 0.49 & -     & -   \\
 $10^{12}$ & 0.02 & 30 & -     & -    & 0.04  & 2.5 \\
 $10^{13}$ & 0.02 & 30 & -     & -    & 0.04  & 0.09 \\
 $10^{14}$ & 0.02 & 30 & -     & -    & 0.04  & 0.004 \\
 $10^{12}$ & 0.02 & 45 & -     & -    & 0.028  & 1.4 \\
 $10^{13}$ & 0.02 & 45 & -     & -    & 0.028  & 0.05 \\
 $10^{14}$ & 0.02 & 45 & -     & -    & 0.028  & 0.002 \\
 $10^{12}$ & 0.02 & 60 & -     & -    & 0.023  & 0.8 \\
 $10^{13}$ & 0.02 & 60 & -     & -    & 0.023  & 0.03 \\
 $10^{14}$ & 0.02 & 60 & -     & -    & 0.023  & 0.001 \\
 $10^{12}$ & 0.5  & 0.4& 1     & 16   & -     & -    \\
 $10^{13}$ & 0.5  & 0.3& 3.35  & 3.2  & -     & -    \\
 $10^{14}$ & 0.5  & 0.1& 11.24 & 0.42 & -     & -    \\
 $10^{12}$ & 0.5  & 30 & 0.84  & 18   & -     & -    \\
 $10^{13}$ & 0.5  & 30 & -     & -    & 1.00  & 2.2  \\
 $10^{14}$ & 0.5  & 30 & -     & -    & 1.00  & 0.08 \\
 $10^{12}$ & 0.5  & 45 & 0.68  & 17   & -     & -    \\
 $10^{13}$ & 0.5  & 45 & -     & -    & 0.707  & 1.3 \\
 $10^{14}$ & 0.5  & 45 & -     & -    & 0.707  & 0.05 \\
 $10^{12}$ & 0.5  & 60 & 0.571 & 13   & -      & -    \\
 $10^{13}$ & 0.5  & 60 & -     & -    & 0.577  & 0.8 \\
 $10^{14}$ & 0.5  & 60 & -     & -    & 0.577  & 0.03 \\
\hline
\end{tabular}
\end{minipage}
\end{table*}

In this subsection we estimate ages of radio pulsars assuming that
the energy losses of an isolated neutron star are associated with
longitudinal currents in the pulsar magnetosphere. In our
estimations  we take into account the evolution of an angle between
the magnetic and the spin axes of a neutron star. The energy losses
of an isolated neutron star estimated by magnetodipole formula are
of the same order of magnitude as the so-called current losses --
the energy release determined by ponderomotive forces of
longitudinal electric currents, i.e. the currents flowing along the
open magnetic field lines and closing on the surface of the star
\citep{bgi1993}.

A simple age estimate of a normal radio pulsar for the model based
on current losses is the same as for the magnetodipole model:
$\tau_{\mathrm{ch}}=P/ 2 \dot{P} \approx 10^7$~yrs. A more accurate
estimate of a radio pulsar lifetime for the current losses model
requires to consider evolution of an angle $\chi$ between the spin
and the magnetic axes of a neutron star.
 The deceleration of neutron star rotation in the scope of this model
 is caused by the surface currents that flow across the magnetic field
 lines \citep{bgi1993}. Therefore, the ponderomotive Ampere force arises,
 and the direction of the appropriate deceleration torque is opposite to
 the magnetodipole moment. Thus, the component of the angular velocity
 of a neutron star $\Omega \sin \chi$ perpendicular to the magnetodipole
 moment (and so perpendicular to the deceleration moment) remains constant, and the ratio

\begin{equation}
 I_{\mathrm{cl}}=\frac{\sin \chi}{P}
\label{Invcl}
\end{equation}
is invariant during the stage of normal radio pulsar.
As we can see, in the scope of current losses model
the angle $\chi$ between the magnetic and the
spin axes evolves to $90^{\circ}$.
A time of evolution of an inclination angle is of the same
order of magnitude as the slow-down rate of a neutron star.

\citet{bgi1993} developed a theory in which energy losses are
associated with longitudinal currents in the neutron star
magnetosphere. In this approach there is no unique formula for
$\dot P$ for all phases of a radio pulsar evolution, also
different assumptions about particle escape can be made. For the
escape model based on the proposal by \cite{rs1975} (hindered
escape) and for pulsars not very close to the death line, the
following equation is derived:

\begin{equation}
\dot P=10^{-15}B_{12}^{10/7}P^{1/14}\cos^{3/2} \chi.
\label{dotPcl}
\end{equation}
Note, that the exact value of the coefficient in this equation is not well
defined. In this paper we use the value written above.
So,  in the scope of the current losses model, we can write the following
expression for the time that an isolated neutron star lives as a radio
pulsar:

\begin{equation}
 \tau_{\mathrm{cl}}[{\mathrm s}]=\int_{P_0}^{P_{\mathrm{tr}}}\frac{dP}{10^{-15}B_{12}^{10/7}P^{1/14}
\left(1- \sin^2 \chi_0(P^2)/(P_0^2) \right)^{0.75}}.
\label{taucl}
\end{equation}
$P_0$ and $\chi_0$ are the initial spin period and the initial angle between the spin
and the magnetic axes of a radio pulsar, respectively,
and $P_{\mathrm{tr}}$ is a period at
which neutron star is leaving the radio pulsar stage.
To find $P_{\mathrm{tr}}$ let us
consider two possible subsequent steps of the evolution of an isolated neutron star.
 At first, as a radio pulsar spins down, it may cross the death line and
pass to the region of the so-called extinct radio pulsars. These neutron
stars have already ceased to radiate in the radio band, but electrodynamic
processes in their magnetospheres still play a crucial role.
In this case the
critical period $P_{\mathrm{dl}}$ is defined by the death line of radio
pulsars:

\begin{equation}
P_{\mathrm{dl}}=B_{12}^{8/15}(\cos \chi)^{0.3} \, \mathrm{s}.
\label{deathlinecl}
\end{equation}
Using  Eq.~(\ref{Invcl}),
we can find an equation which shows
how the period of transition of a neutron star
to the region of extinct radio pulsars depends on the initial parameters:

\begin{equation}
P_{\mathrm{dl}}=B_{12}^{8/15}(1-\sin^2 \chi_0 \frac{P_{\mathrm{dl}}^2}{P_0^2})^{0.15}.
\label{Pdlcl}
\end{equation}

If the model based on the current losses is valid, the period
derivative (\ref{dotPcl}) reduces during the evolution of a
neutron star due to the factor $\cos^{3/2} \chi$. So, the longest
part of the lifetime of an active radio pulsar is the time it
spends near the death line (unless a radio pulsar passes to the
stage of orthogonal rotators before it can reach the death line -–
this way of evolution we discuss below). Thus, the age estimate
(\ref{taucl}) significantly depends  on the accuracy of the
expression (\ref{dotPcl}) near the death line. Unfortunately, as
is was noted above, expression (\ref{dotPcl}) adequately defines
the period derivative only for the radio pulsars far from the
death line. Thus, the age estimates for the current losses model
are very approximate and the more accurate expression for the
spin-down rate of a neutron star near the death line can alter the
results obtained in (\ref{taucl}) by a factor of a few.

%In addition we want to pay attention to another possibility for a
%neutron star to leave the region of bright radio pulsars.
As it is noted in the previous paragraph, a NS instead of crossing
the death line can follow a different path. An alternative
evolutionary track leads an object directly to the stage of
orthogonal rotators \citep{bgi1993, bn2004}.
 Here a short explanation should be done.
If an inclination angle of a radio pulsar is
close enough to $90^{\circ}$, so that

\begin{equation}
\left| \pi/2 - \chi \right| < R_0/R,
\label{ortbound}
\end{equation}
then particles with opposite charge signs can flow out of the polar
cap surface. Here $R_0= R \left( \Omega R/c \right)^{0.5}$ is the
polar cap radius. A group of radio pulsars for which the
Goldreich-Julian charge density changes the sign on the polar cap
surface is called orthogonal rotators. These pulsars are
characterized by fast ($P \sim 0.3$~s) or superfast ($P < 0.1$~s)
rotation, reduced spin-down rate ($\dot P \sim 10^{-17}-10^{-19}$
s~s$^{-1}$), weak radio emission ($L_{\mathrm{rad}} \sim
10^{25}-10^{27}$~erg~s$^{-1}$), and the existence of an interpulse.
The energy losses for this class of objects are reduced by the
factor $\Omega R/c$ in comparison with the normal radio pulsars and
are virtually independent of the angle between the magnetic and the
spin axes, which is anyway very close to 90 degrees. Some candidates
to this class are listed in \citep{bn2004}. They have periods from
$\sim 100$ up to $\sim 500$ msec. Not all radio pulsar with small
$\dot P$ should be orthogonal rotators. It was noted by
\cite{lor2004} that isolated pulsars with small period derivatives,
$\sim 10^{-18}$, and  shorter periods about few dozens of
milliseconds can be ''disrupted recycled pulsars'', but objects with
longer spin periods require a different explanation which is
probably related to emission mechanism.

The boundary of the stage of orthogonal rotators is given by
(\ref{ortbound}):

\begin{equation}
\cos \chi=\left( 2\pi R / cP \right)^{1/2}=1.45 \times
10^{-2}P^{-0.5}, \label{ortcl}
\end{equation}
so it is easy to find the following implicit equation for the
period of the transition to this stage:

\begin{equation}
P_{\mathrm{ort}}= \frac{P_0}{\sin \chi_0} \left(1-2.1 \times
10^{-4}P_{\mathrm{ort}}^{-1} \right)^{0.5}.
\end{equation}

The death line of radio pulsars (\ref{deathlinecl}) and the boundary of the
orthogonal rotators stage (\ref{ortcl}) intersect at the point
$P_{\mathrm{int}} \approx 0.35 B_{12}^{0.46}$ s, $\sin \chi_{\mathrm{int}}
\approx \left( 1-6 \times 10^{-4}B_{12}^{0.46} \right)^{0.5}$ in the $P$ --
$\sin \chi$ diagram. Therefore, if

\begin{equation}
\sin \chi_{\mathrm{0}}/P_{\mathrm{0}} >
\frac{(1-6 \times 10^{-4}B_{12}^{0.46})^{0.5}}{0.35 B_{12}^{0.46}},
\label{sincl}
\end{equation}
a neutron star passes through the death line to the region of
extinct radio pulsars. In this case, the period of transition is
determined by the death line (\ref{deathlinecl}), so that
$P_{\mathrm{tr}}=P_{\mathrm{dl}}$. Thus, assuming the initial spin
period $P_0=20$~ms, the initial angle between the magnetic and the
spin axes $\chi_0=1^{\circ}$, and the magnetic field $B=10^{12}$~G,
from Eq.~(\ref{taucl})  we can estimate the age of a radio pulsar at
the point of the transition to the region of extinct radio pulsars
as $\approx 36$~Myr. The spin period of a neutron star at the point
of the transition is $P_{\mathrm{dl}} \approx 0.88$~s, and the
inclination angle is $\chi_{\mathrm{dl}} \approx 50^{\circ}$. For
the same spin period and standard period derivative $\dot
P=10^{-15}$ $\mathrm{ss}^{-1}$, the characteristic age estimate
$\tau_{\mathrm{ch}}=P/ 2 \dot P$ results in $14$~Myr.
%If the initial
%parameters (the initial spin period and inclination angle) are the
%same and the magnetic field is $B=10^{13}$~G then Eq.~(\ref{taucl})
%gives us $\tau_{\mathrm{cl}} \approx 3$~Myr ($P_{\mathrm{dl}}
%\approx 1.145$~s, $\chi_{\mathrm{dl}} \approx 87.6^{\circ}$).
As we can see, taking into account the evolution of the angle
between the spin and the magnetic axes within the current losses
model significantly increases the time that a radio pulsar needs to
slow down from the spin period $P_1$ to the larger spin period $P_2$
(both $P_1$ and $P_2$ are supposed to belong to the region of active
radio pulsars).

Otherwise, if the initial parameters of a pulsar do not meet the
above condition (\ref{sincl}), the transition to the stage of
orthogonal rotators takes place, so that
$P_{\mathrm{tr}}=P_{\mathrm{ort}}$. Note, that the transition to the
orthogonal rotator stage should be typical for most of the radio
pulsars with comparatively short initial spin periods $P_0 \sim
10$~ms (the ratio $\sin \chi/P$ is initially very large for these
radio pulsars provided that the angle $\chi$ is not too close to
$0^{\circ}$). Thus, for the model based on current losses the
majority of neutron stars pass to the orthogonal rotator stage
before the transition to the region of extinct radio pulsars.
Assuming the same initial period $P_0=20$~ms and
$\chi_0=45^{\circ}$, we can estimate the age of a neutron star
passing from the stage of radio pulsars to the orthogonal rotator
stage as follows: $\tau_{\mathrm{ort}}=1.44$~Myr (assuming
$B=10^{12}$~G), $\tau_{\mathrm{ort}}=54000$~yrs ($B=10^{13}$~G), and
$\tau_{\mathrm{ort}}=2000$~yrs ($B=10^{14}$~G). The spin period at
which a radio pulsar crosses the boundary of the orthogonal rotator
stage for all three magnetic field strengths is the same
$P_{\mathrm{ort}} \approx 28.2$~ms. As one can see, it requires
about 1~Myr for an active radio pulsar with comparatively large
initial angle $\chi$ and standard magnetic field to increase its
period by only 8~ms. Table~2 contains the age estimates for
different parameters (such as magnetic fields, initial spin periods
and initial inclination angles of radio pulsars).

Thus, we show that if the evolution of an inclination angle $\chi$
is taken into account, the slow-down rate of an active radio pulsar
turns out to be several times lower, and a lifetime of a star at
that stage turns out to be much longer than it is commonly assumed.
Moreover, as the angle $\chi$ between the spin and the magnetic axes
of a pulsar increases, the slow-down rate of the star reduces by the
factor of $(\cos \chi)^m$, where the power $m$ is $\approx
1.5$~--~2. Furthermore, as it is discussed in \citep{bn2004}, a
period derivative $\dot P$ of neutron stars at the stage of
orthogonal rotators becomes approximately $10^4$ smaller than for
the normal radio pulsars, so we should see some weakly radiating old
radio pulsars with comparatively small spin periods and with the
angles between the magnetic and the spin axes close enough to
$90^{\circ}$.

After all, if radio pulsars are born as fast rotators and their
initial angle distribution is equiprobable, for the model of current
losses the most of radio pulsars should become orthogonal rotators
long before they can pass through the death line. Thus, as opposed
to the magnetodipole model, we can expect the distribution in
present day angles between the magnetic and the spin axes of active
radio pulsars to have its maximum close to $90^{\circ}$. If the
angle distribution has no such peak, we can assume that either the
initial inclination angles are very small, or an essential part of
observable radio pulsars was born with long spin periods. The latter
conclusion is quite probable, and has a long history of discussion
in the literature (see, for example, \citet{hl2006} and references
therein).

\section{Reconstruction of evolutionary tracks: current losses model}
\label{appendix}

As it is shown above, the evolution of the inclination angle of an
isolated neutron star significantly affects the estimate of its
characteristic age through the influence of this parameter on the
spin-down rate and the shape of evolutionary track of a neutron
star. Usually, evolutionary tracks of NSs are shown on the
$P$~--~$\dot P$ plot. However, in the case of evolving $\chi$ it
is convenient to use the $P$~--~$\sin \chi$ diagram. In this
section we describe the procedure of reconstruction of
evolutionary tracks for the isolated neutron stars assuming that
the model based on current losses is valid. Two models of particle
acceleration region should be considered: the model, in which the
longitudinal electric field near the neutron-star surface hinder
the particle escape \citep{rs1975}
%(the so-called model with hindered particle escape, \citealt{rs1975})
and the model, in which the longitudinal electric field does not
exist near the neutron-star surface (the model with free particle
escape, \citealt{a1979}; \citealt{m1999}).

The evolutionary track of a neutron star in the $P-\sin \chi$
diagram does not depend on the magnetic field strength directly.
At that, the rate at which a neutron star moves along its
evolutionary track depends on the magnetic field. Besides, the
magnetic field strength affects the death line and the line of the
transition to the propeller stage, so it can affect the
evolutionary track through the track lengths of a neutron star at
different stages of evolution. Further, we want to note that the
boundary of the region of orthogonal rotators does not depend on
the magnetic field either. At last, we should emphasize that the
bias of the track in the region of active radio pulsars is defined
solely by the ratio (\ref{Invcl}) for the current losses model.
Therefore, while constructing the evolutionary tracks by the
initial parameters (direct way), one should keep in mind that Eq.~
(\ref{Invcl}) is very susceptible to the small changes in the
initial angle value in the case of the short (millisecond) initial
period.

%Once again we want to emphasize that all age estimates of the
%neutron stars obtained in the previous sections are valid only
%under the assumption that presently observed parameters of a
%neutron star still belong to the region of radio pulsars in the
%$P$--$\sin \chi$ plane. As one can see,
Many of the newly discovered types of isolated neutron stars have
long spin periods and may belong to the next evolutionary stages
such as extinct radio pulsars or supersonic propellers, or even to
orthogonal or aligned rotators. In that case, in order to estimate
the age of the star we should reconstruct the whole evolutionary
track of a neutron star and consider a neutron star age as a sum
of time intervals during which a star lives at each of the stages.

As example, in this section we present reconstructions of
evolutionary tracks for two NSs belonging to different classes:
the Magnificent seven and anomalous X-ray pulsars.
%the following two neutron stars: X-ray dim
%isolated neutron star RX~J0720.4-3125 and AXP~1E2259+586.
The choice is motivated at first by the fact that for both of them
some information about inclinations angles is available. For one
of these NSs (RX~J0720.4-3125) the angle is small or intermediate,
as it appears from modeling it's X-ray light curve (see below).
For another, AXP~1E2259+586, the angle is large, as it is clear
from the double peaked X-ray pulse profile (see, for example,
\citet{wt2004}). Then both of these NSs has relatively high
magnetic fields, which means that the rate of their
magneto-rotational evolution is high, too. Finally, poor, but
independent, age estimates are available for them.  In particular,
the characteristic age for RX~J0720.4-3125 is $\tau = P/2\dot P=
1.9$~Myr \citep{h2006} is  larger than it can be estimated from
cooling curves. Cooling age should be $<$~1~Myr \citep{page2004}.
In each of the following plots an independent age estimate is
marked by an asterisk. The independent age estimates are taken as
1~Myr for RX J0720.4-3125 (excluding two plots with the
characteristic ages less than the independent ones, in which cases
the mark is placed at the origin, indicating that initial period
and angle should be small) and 0.01~Myr for 1E 2259+586. The
latter is based on the associations of the source with supernova
remnant CTB~109 \citep{kg2003}.

%We select inclination angles and magnetic field strengths so that
%these neutron stars can belong either to the region of active
%radio pulsars or to the region of extinct radio pulsars. Thus, the
%possibility for a neutron star to be positioned on the $P$--$\sin
%\chi$ diagram within the stages of orthogonal rotators, aligned
%rotators or supersonic propeller is not considered in this
%section. Observational data support this assumption.

At first, we consider the evolution of the X-ray dim isolated
neutron star RX~J0720.4-3125 within the model of current losses,
and reconstruct possible evolutionary tracks in $P$~--~$\sin \chi$
plane assuming that the angle between spin and magnetic axes at
the present time is $\chi_{\mathrm{obs}}=50^{\circ}$
\citep{ppm2006}. Basing on the present day period
$P_{\mathrm{obs}}=8.39$~s and the angle $\chi_{\mathrm{obs}}$,
this object can be either in the region of active radio pulsars or
in the region of extinct radio pulsars depending on the magnetic
field strength. We consider the
 reverse track reconstruction, i.e. knowing the
parameters (the spin period and the inclination angle) of the
neutron star at the present time we reconstruct the evolutionary
track. After that we can find, how the initial parameters are
related to each other. The initial spin period $P_0=20$~ms is used
only for the age estimate of the neutron star.

Here we reconstruct the evolutionary tracks of RX~J0720.4-3125
assuming the following three values of the magnetic field:
$B=10^{14}$~G, $B=10^{13.5}$~G, and $B=10^{13}$~G. Knowing the
magnetic field strength, we can define the death line of radio
pulsars and the lines of the transitions to the stages of
orthogonal rotators and supersonic propellers. After that we can
determine if the point with the presently observed spin period
$P_{\mathrm{obs}}$ and the presently observed inclination angle
$\chi_{\mathrm{obs}}$ is located in the region of active radio
pulsars or it falls to the extinct radio pulsar stage. According
to the position of this point on $P$~--~$\sin \chi$ diagram, we
can define the next steps to reconstruct the evolutionary path.

\begin{figure}
 \includegraphics[width=0.35\textwidth, keepaspectratio,angle=-90]{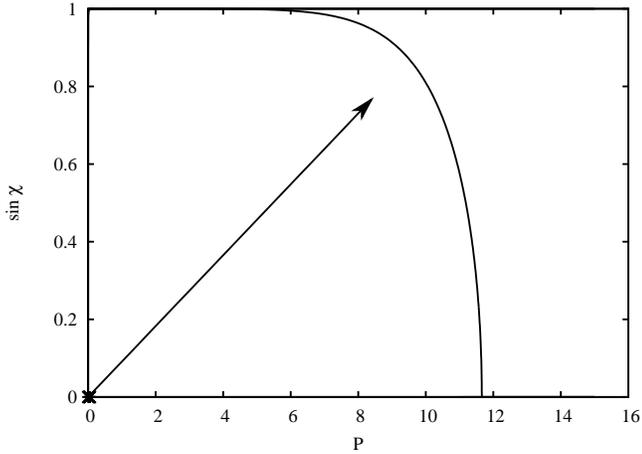}
\caption{ The reconstructed evolutionary track of XDIN
J0720.4-3125. $B=10^{14}$~G, $P_{\mathrm{obs}}=8.39$~s,
$\chi_{\mathrm{obs}}=50^{\circ}$. Model based on current losses is
used. The figure is valid for both hindered and free particle
escape. The presently observed spin period and inclination angle
correspond to the region of active radio pulsars. If the initial
spin period $P_0=20$~ms, the age estimate is $\tau \approx
0.4$~Myr. The ''star'' marks the initial parameters correspondent
to the independent age estimate $\tau_{\mathrm ind}$=0.4~Myr.}
\label{0720c50B14}
\end{figure}

If we assume the magnetic field $B=10^{14}$~G, the present
parameters of J0720.4-3125 correspond to the point in the region
of active radio pulsars. Recall, that during this stage the ratio
$\sin \chi/P$ is invariant. Therefore, the evolutionary track of
RX J0720.4-3125 can be represented by a straight line between the
points  ($P_{\mathrm{obs}}; \sin \chi_{\mathrm{obs}}$) and  (0;0)
(see Fig.~\ref{0720c50B14}). The point defining the initial spin
period $P_0$ and the initial inclination angle $\chi_0$ is located
on this line; thus, if we know the initial spin period, we can
unambiguously define the initial angle between the magnetic and
the spin axes. If the initial spin period is short enough
($P_0=20$~ms), the age of RX J0720.4-3125 can be estimated as
0.41~Myr (\ref{taucl}).

%The characteristic age for this NSs is $\tau = P/2\dot P= 3.8$~Myr
%\citep{h2006}, that is much higher than it can be estimated from
%cooling curves. Cooling age should be $<$~1~Myr \citep{page2004}.

%\renewcommand{\thefigure}{\arabic{figure}\alph{subfigure}}
%\setcounter{subfigure}{1}

For lower magnetic field strengths the observed parameters of RX
J0720.4-3125 correspond to the region of extinct radio pulsars. In that
case, the evolutionary tracks of the isolated neutron star will be different
for the following
two models of an acceleration region: the model with hindered particle
escape \citep{rs1975} and the model with free particle escape
\citep{a1979,m1999} from the neutron star surface.

Within the model with hindered particle escape plasma fills only the
inner region of the pulsar magnetosphere, and a neutron star slows
down according to the magnetodipole formula (\ref{dotPmd}),
therefore $\cos \chi /P$ is invariant for extinct radio pulsars.
Hence, the track for this XDIN in the region of extinct radio
pulsars in $P - \sin \chi$ plane can be represented as follows:

\begin{equation}
\sin \chi = (1-P^2 \left( \cos \chi_{\mathrm{obs}} /P_{\mathrm{obs}} \right)^2)^{0.5}.
\label{cosext}
\end{equation}

In order to continue the evolutionary track of the star to the region of
active radio pulsars we should find the parameters, at which neutron star
passes through the death line of radio pulsars (\ref{Pdlcl}). As a
result, for RX J0720.4-3125 we have:

\begin{equation}
P_{\mathrm{dl}}=B_{12}^{0.75}
\left( \frac{\cos \chi_{\mathrm{obs}}}{P_{\mathrm{obs}}}
\right)^{0.41} \approx 0.35 B_{12}^{0.75} \, {\mathrm s},
\end{equation}

\begin{equation}
\cos \chi_{\mathrm{dl}}=B_{12}^{0.75}
\left( \frac{\cos \chi_{\mathrm{obs}}}{P_{\mathrm{obs}}} \right)^{1.41}
\approx 0.027 B_{12}^{0.75}.
\end{equation}

Knowing the coordinates of the point, where the evolutionary track
for the extinct radio pulsar region intersects the death line, we
can continue the evolutionary track of RX 0720.04-3125 in the region
of active radio pulsars (the procedure is the same as it is shown
for the example with $B=10^{14}$~G).

According to (\ref{dotPmd}), by the time of observation the neutron star
lives as an extinct radio pulsar:

\begin{equation}
\tau_2=\int_{P_{\mathrm{dl}}}^{P_{\mathrm{obs}}}\frac{dP}{2.4 \times 10^{-16}B_{12}^{2}P^{-1} \left
(1-\cos^2 \chi_{\mathrm{obs}}(P^2)/(P_{\mathrm{obs}}^2) \right)} \, {\mathrm s}.
\end{equation}

Finally, for the lifetime of a neutron star as an extinct radio pulsar at the time
of observation we have:

\begin{eqnarray}
\tau_2=2 \times 10^{15} B_{12}^{-2}P_{\mathrm{obs}}^2 \cos^{-2}
\chi_{\mathrm{obs}} \times
\\
\nonumber
\times \, \ln \left|
\frac{P_1^2-P_{\mathrm{obs}}^2/\cos^2
\chi_{\mathrm{obs}}}{P_{\mathrm{obs}}^2-P_{\mathrm{obs}}^2/\cos^2
\chi_{\mathrm{obs}}} \right| \, {\mathrm s}.
\end{eqnarray}

Thus, the age of an extinct radio pulsar can be estimated as:

\begin{equation}
\tau=\tau_1+\tau_2,
\end{equation}
where $\tau_1$ is the time neutron star lives as radio pulsar
(\ref{taucl}) and $\tau_2$ is a lifetime in the region of extinct
radio pulsars.

%In each plot we mark by a cross independent age estimates, which
%are taken as 1~Myr for RX J0720.4-3125, and 0.01~Myr -- for 1E
%2259+586. The former estimate is a rough upper limit based on
%cooling curves \citep{page2004}. The latter is based on the
%associations of the source with supernova remnant CTB~109
%\citep{kg2003}.

\begin{figure}
\includegraphics[width=0.35\textwidth, keepaspectratio,angle=-90]{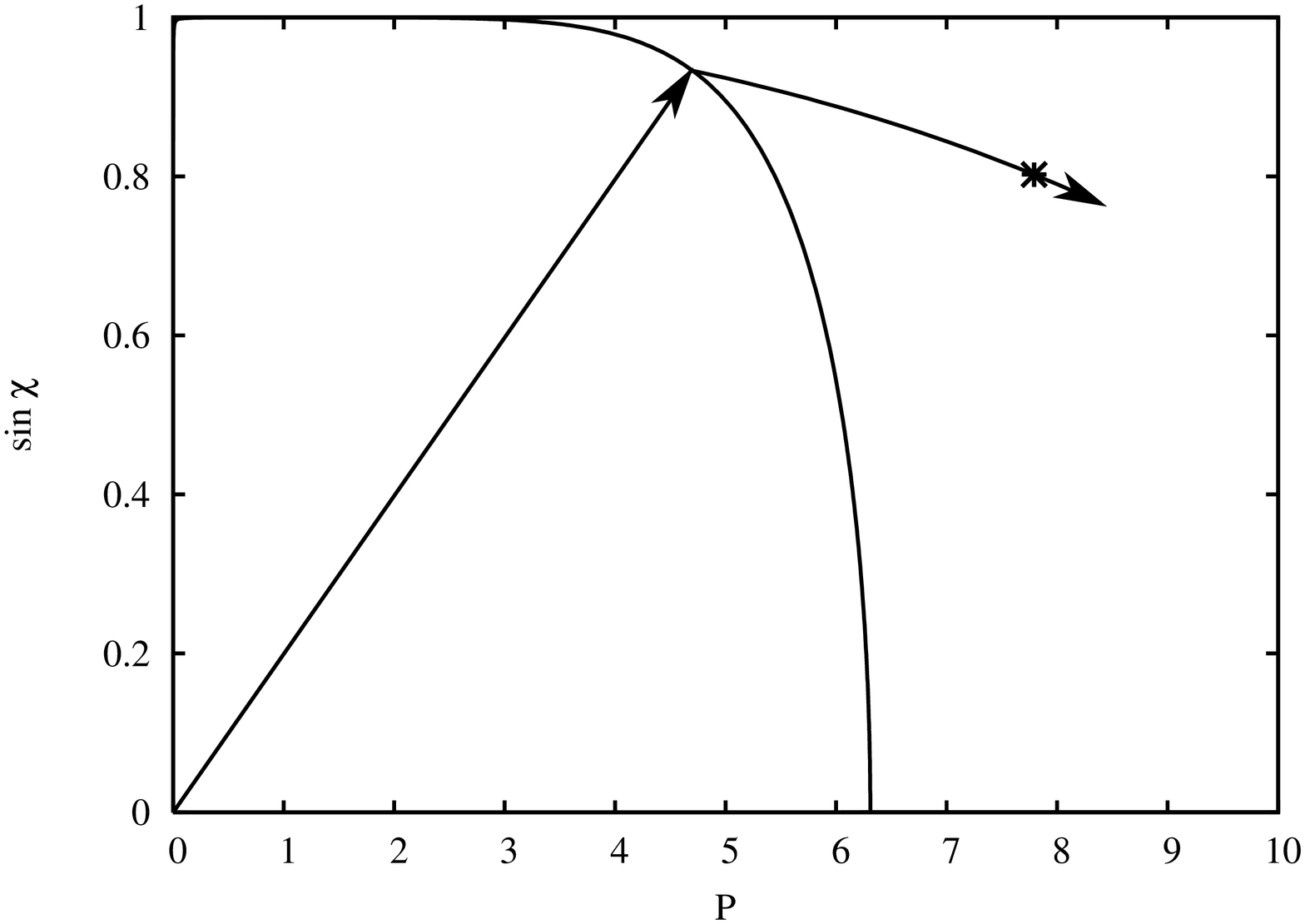}
 \caption{The reconstructed evolutionary track of XDIN J0720.4-3125.
 $B=10^{13.5}$~G, $P_{\mathrm{obs}}=8.39$~s, $\chi_{\mathrm{obs}}=50^{\circ}$.
 Model based on current losses is used. The figure is valid for hindered particle escape.
 The presently observed spin period and inclination angle correspond to the region
 of extinct radio pulsars. If the initial spin period $P_0=20$~ms,
 the lifetime at the active radio pulsar stage is $\tau_1 \approx 1.5$~Myr,
 and the lifetime at the extinct radio pulsar stage is $\tau_2 \approx 4.3$~Myr.
 The total age estimate is $\tau \approx 5.8$~Myr.
 The asterisk marks the initial parameters correspondent to the
independent age estimate $\tau_{\mathrm {ind}}$=1~Myr.}
\label{0720c50B135}
\end{figure}

Fig.~\ref{0720c50B135} shows the reconstructed evolutionary track
for RX~J0720.4-3125 within the model with hindered particle escape
from the neutron star surface for magnetic field strengths
$B=10^{13.5}$~G.

If the model with free particle escape from the neutron star surface is
valid then an angle between the spin and the magnetic axes tends to
$90^{\circ}$ for both active and extinct radio pulsar regions. So, we can
simply follow the same procedure for the extinct radio pulsars as we define
for the active radio pulsars (see Fig.~\ref{0720c50B14} and the example with
$B=10^{14}$~G above) and use (\ref{taucl}). Assuming the spin period of the
star at birth was $P_0=20$~ms, the age of X-ray dim isolated neutron star RX
J0720.4-3125 within the model with free particle escape can be estimated as
11~Myr and 2~Myr for $B=10^{13}$~G and $B=10^{13.5}$~G, respectively.
It should be noted here, that the theory of the energy release associated
with the longitudinal currents in the neutron star magnetosphere was
generally developed by \cite{bgi1993} under the assumption of the hindered
particle escape \citep{rs1975}. Thus, if particles leave the surface of a
neutron star freely (\citealt{a1979}; \citealt{m1999}), expression
(\ref{dotPcl}) is a rough approximation of the period derivative and so the
ages estimated under the assumption of the free particle escape can be
altered by a factor of a few for the more accurate spin-down rate
expression.

In addition, in Figs. \ref{2259c80B613}--\ref{2259c85B613} we
present the evolutionary tracks of AXP~1E2259+586 for a set of
possible values of magnetic field strength and presently observed
angle between the magnetic and the spin axes (see data on AXPs in
general and 1E2259+586 in particular in \citet{wt2004}). Finally,
in Figs. \ref{0720c5B135}--\ref{0720c5B14} we reconstruct the
evolutionary tracks of RX~0720.04-3125, assuming
$\chi_{\mathrm{obs}}=5^{\circ}$ \citep{haberl2006}. The validity
of the model with hindered particle escape from the neutron star
surface is assumed in these figures.

As, most probably, both objects left the region of active radio
pulsars, then the track reconstruction shows that initial periods
of these objects are long enough.

%\addtocounter{figure}{-1}
%\addtocounter{subfigure}{1}

%\addtocounter{figure}{-1}
%\addtocounter{subfigure}{1}

\renewcommand{\thefigure}{\arabic{figure}}
\begin{figure}
 \includegraphics[width=0.35\textwidth, keepaspectratio,angle=-90]{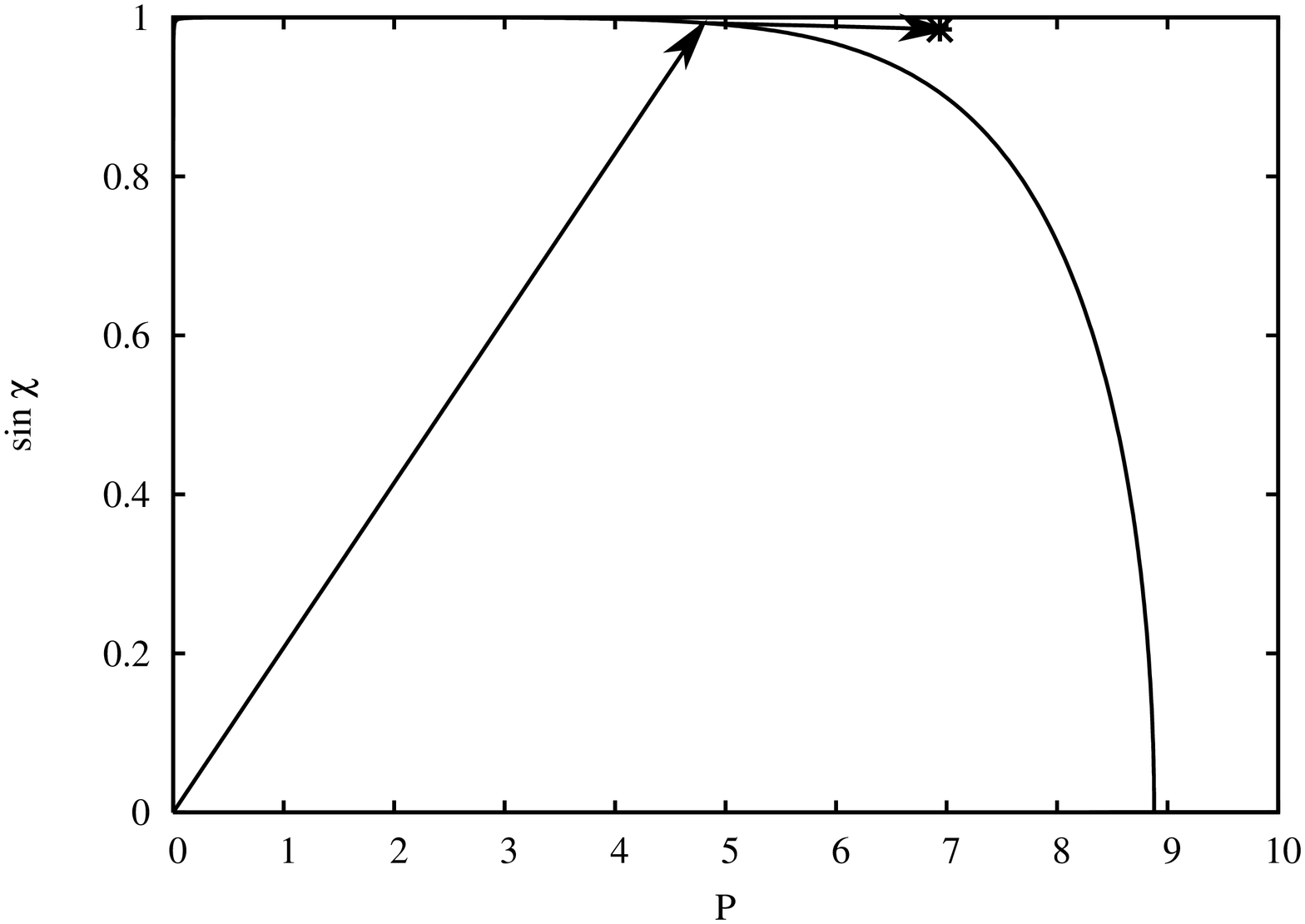}
\caption{The reconstructed evolutionary track of AXP 1E2259+586.
$B=6 \times 10^{13}$~G, $P_{\mathrm{obs}}=6.98$~s,
$\chi_{\mathrm{obs}}=80^{\circ}$. Model based on current losses is
used. The figure is valid for hindered particle escape. The
presently observed spin period and inclination angle correspond to
the region of extinct radio pulsars. If the initial spin period
$P_0=20$~ms, the lifetime at the active radio pulsar stage is
$\tau_1 \approx 0.8$~Myr, and the lifetime at the extinct radio
pulsar stage is $\tau_2 \approx 0.5$~Myr. The total age estimate
is $\tau \approx 1.3$~Myr. The asterisk marks the initial
parameters correspondent to the independent age estimate
$\tau_{\mathrm {ind}}$=0.01~Myr.} \label{2259c80B613}
\end{figure}

%\addtocounter{figure}{-1}
%\addtocounter{subfigure}{1}

\begin{figure}
 \includegraphics[width=0.35\textwidth, keepaspectratio,angle=-90]{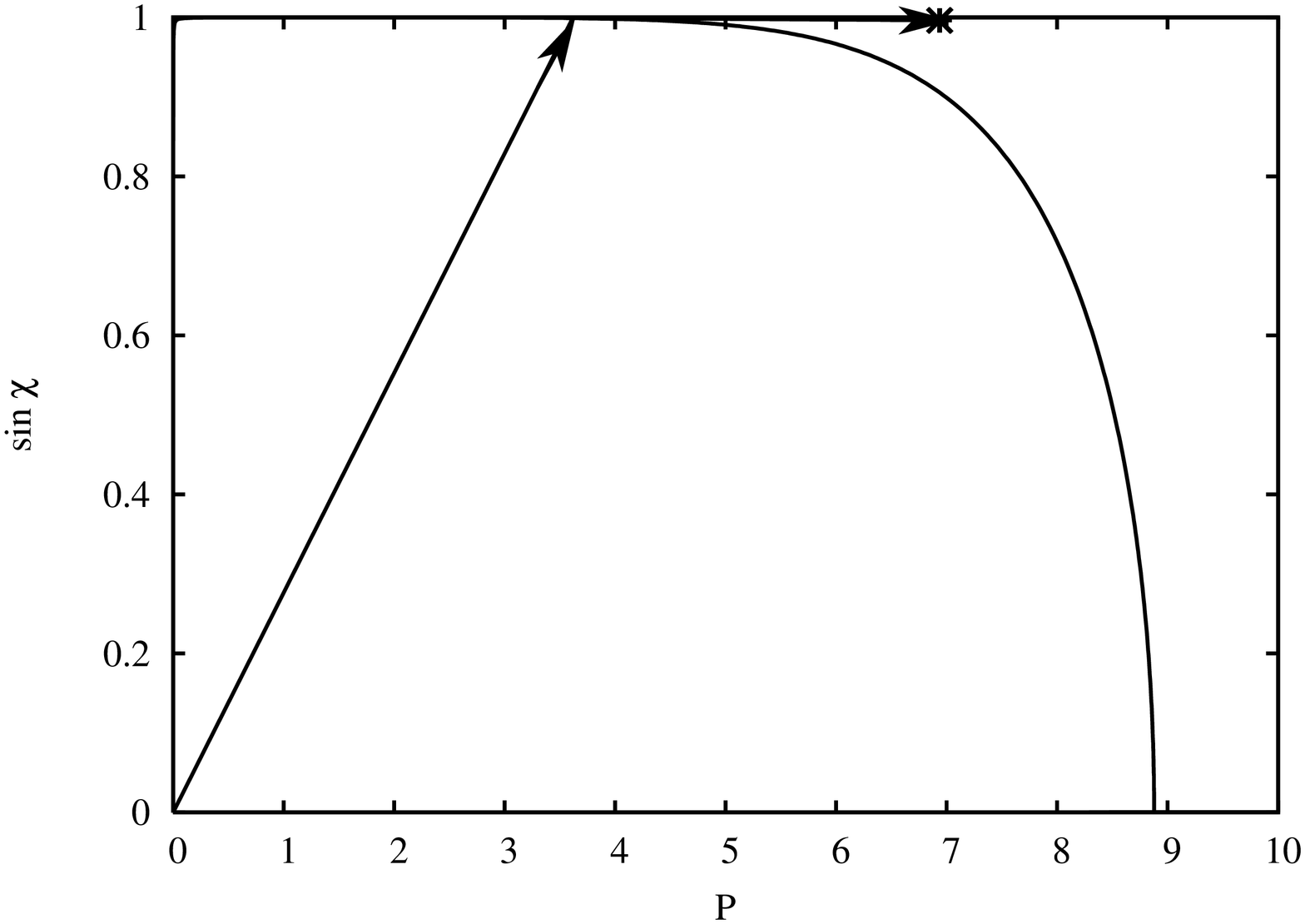}
\caption{The reconstructed evolutionary track of AXP 1E2259+586.
$B=6 \times 10^{13}$~G, $P_{\mathrm{obs}}=6.98$~s,
$\chi_{\mathrm{obs}}=85^{\circ}$. Model based on current losses is
used. The figure is valid for hindered particle escape. The
presently observed spin period and inclination angle correspond to
the region of extinct radio pulsars. If the initial spin period
$P_0=20$~ms, the lifetime at the active radio pulsar stage is
$\tau_1 \approx 0.7$~Myr, and the lifetime at the extinct radio
pulsar stage is $\tau_2 \approx 0.6$~Myr. The total age estimate
is $\tau \approx 1.3$~Myr. The asterisk marks the initial
parameters correspondent to the independent age estimate
$\tau_{\mathrm {ind}}$=0.01~Myr.} \label{2259c85B613}
\end{figure}

%\addtocounter{figure}{-1}
%\addtocounter{subfigure}{1}

\begin{figure}
 \includegraphics[width=0.35\textwidth, keepaspectratio,angle=-90]{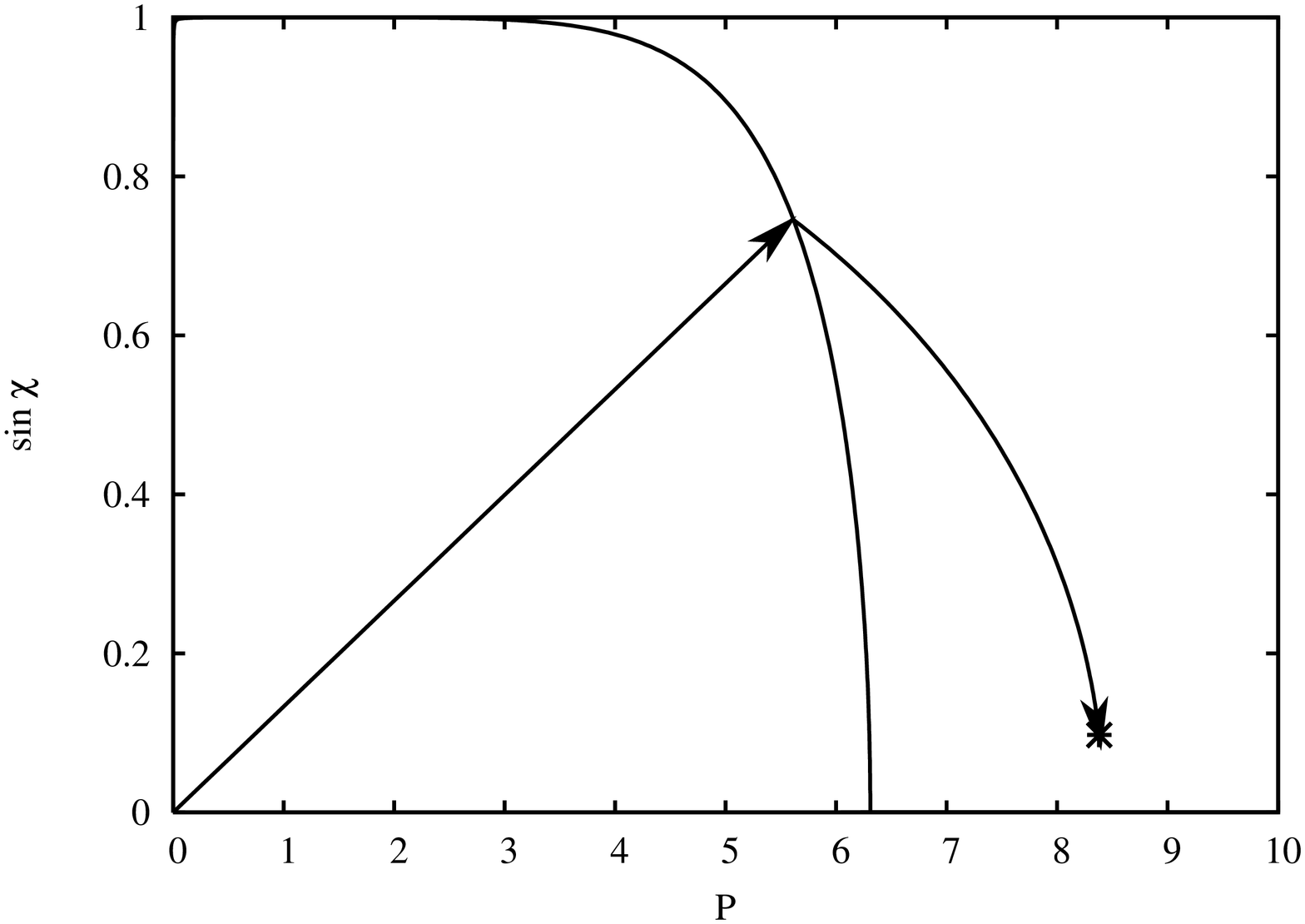}
\caption{The reconstructed evolutionary track of XDIN
J0720.4-3125. $B=10^{13.5}$~G, $P_{\mathrm{obs}}=8.39$~s,
$\chi_{\mathrm{obs}}=5^{\circ}$. Model based on current losses is
used. The figure is valid for hindered particle escape. The
presently observed spin period and inclination angle correspond to
the region of extinct radio pulsars. If the initial spin period
$P_0=20$~ms, the lifetime at the radio pulsar stage is $\tau_1
\approx 1.4$~Myr, and the lifetime at the extinct radio pulsar
stage is $\tau_2 \approx 19.3$~Myr. The total age estimate is
$\tau \approx 20.7$~Myr. The asterisk marks the initial parameters
correspondent to the independent age estimate $\tau_{\mathrm
{ind}}$=1~Myr.} \label{0720c5B135}
\end{figure}

%\addtocounter{figure}{-1}
%\addtocounter{subfigure}{1}

\begin{figure}
 \includegraphics[width=0.35\textwidth, keepaspectratio,angle=-90]{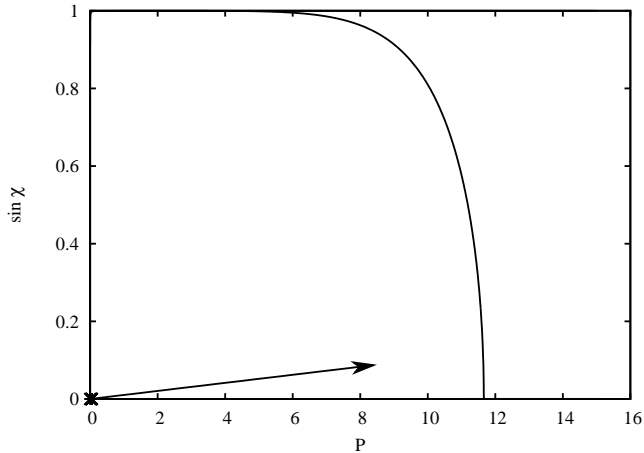}
\caption{The reconstructed evolutionary track of XDIN
J0720.4-3125. $B=10^{14}$~G, $P_{\mathrm{obs}}=8.39$~s,
$\chi_{\mathrm{obs}}=5^{\circ}$. Model based on current losses is
used. The figure is valid for both hindered and free particle
escape. The presently observed spin period and inclination angle
correspond to the region of active radio pulsars. If the initial
spin period $P_0=20$~ms, the age estimate is $\tau \approx
0.34$~Myr. The asterisk marks the initial parameters correspondent
to the independent age estimate $\tau_{\mathrm {ind}}$=0.34~Myr.}
\label{0720c5B14}
\end{figure}

\renewcommand{\thefigure}{\arabic{figure}}

\section{Discussion}

In this section we provide a brief discussion on several topics
related to our analysis.

We discuss possible modifications of models due to
non-electromagnetic effects. Then we discuss the possibility to
choose between free and hindered particle escape basing on the
data on inclinations angles of extinct pulsars.

Then, as the main impact of poor knowledge of $\chi$ evolution is
on age estimates and population synthesis models, we touch these
subjects.
%In particular, we discuss properties of highly
%magnetized NSs.

\subsection{The role of invariants and modifications}

One of highlights of this paper can be summarized as follows. If a
neutron star is born rapidly rotating, it reaches the stage of low
energy losses with very short spin period, unless the initial
inclination angle of a star is very close to the value
correspondent to maximal spin down. This conclusion crucially
depends on the existence of the invariants (\ref{Invmd}) or
(\ref{Invcl}). The existence of the invariants is based on several
simplifying assumptions:

\begin{itemize}
\item A neutron star is an ideal sphere;
\item Magnetifc field is due to an ideal dipole;
\item Magnetic dipole braking or current losses are the only electro-magnetic processes which
slows down the star;
\item There are no additional non electro-magnetic processes related to internal structure of a NS
or to external agents.
\end{itemize}
Clearly, all four assumptions can be violated. In such a case most
of the results presented above have to be reconsidered taking into
account additional processes or/and conditions.

In particular, we want to pay attention to the old paper by
\cite{macy1974}. This author notes, that if the dominant internal
field is poloidal, then due to non-sphericity of a star the
alignment process is working. If, oppositely, the toroidal
component of the field dominates -- then the counteralignment
takes place. This arguments are particularly interesting as
magnetars, especially SGRs, are supposed to have huge toroidal
fields, which are dissipated on the spin-down time scale of these
objects and deform the star \citep{Stella2005}.

Another possibility is related to remnant discs \citep{m2001}. They
can provide additional spin-down without significant changes in the
inclination angle, especially at early stages of evolution. So, when
the disc becomes inefficient a pulsar already has a relatively long
period, but did not significantly change it's inclination angle,
yet.

Also, a process of vortex line migration during spin-down can be
important \citep{r2006}. Effectively, this process results in
counteralignment, and so can partly compensate alignment in
magneto-dipole model. However, as alignment goes as $\cos \chi
\propto P$ and vortex line migration leads to counteralignment as
$\sin \chi \propto P^n,\, 0< n<0.5$, compensation is only partial.

Finally, for a rigid and non-spherical neutron star the free
nutation process plays significant role regulating the tendency of
the magnetic axis to align of counter-align with the spin axis.
\citep{g1970} showed that in case of magnetic dipole braking model
the nutation amplitude decreases if the inclination angle of a
neutron star is less than $\approx 55^{\circ}$ and increases if the
inclination angle is larger than $\approx 55^{\circ}$.

Electromagnetic processes can also lead to small changes in the
angle $\chi$. Probably, it can be the case for PSR B1931+24 which
shows a peculiar behavior \citep{k2006}, if the explanation is in
switching between magneto-dipole and current losses
\citep{bn2006}.

\subsection{Free vs. hindered particle escape}

 Potentially, in the case of the longitudinal current losses
the distribution  in the inclination angle of the neutron stars
behind the death line can provide necessary information to choose
between free \citep{a1979,m1999} and hindered \citep{rs1975}
particle escape. The reasoning is simple: if the model of free
escape is realized, than (especially among the strongly magnetized
neutron stars) most of the neutron stars should have $\chi$ close to
$90^\circ$. In the opposite case, there should be no neutron stars
very close to $\chi=90^\circ$, but this does not mean that most of
them have $\chi\approx 0^\circ$. However, this easy approach can be
done only if neutron stars are significantly evolved, i.e. if their
present-day spin periods are much larger than the initial ones.
Unfortunately, our analysis presented above shows that for the radio
pulsars and magnetars the initial periods cannot be much smaller
than the observed ones if either the magnetodipole or current losses
model is valid.

Among all highly magnetized NSs the Magnificent seven can have
relatively small initial periods in comparison with the present
day values, as it comes out after comparison of there age
estimates based on the standard assumption and its analogue for
current losses with independent estimates. If this is the case,
then it is interesting to note, that they do not show
overabundance of orthogonal rotatotors. This can be considered as
a (weak) argument in favor of the hindered escape model.

\subsection{Age estimates for highly magnetized NSs and their initial spin periods}
\label{ages}

For the current losses model we compare the standard characteristic
ages and the age estimates with the account of the angle evolution
for the different types of isolated neutron stars under the specific
assumption that all considered isolated neutron stars still remain
in the region of active radio pulsars, i.e. they did not pass either
through the death line or through the boundary of the orthogonal
rotator stage. Table~3 presents the results for one individual
neutron star from each of the following types: HB-PSRs, SGRs, AXPs,
X-ray dim isolated neutron stars and RRATs. For some sources  the
validity of the above assumption, of course, can be questioned. But
for the considered objects this assumption results in the lower age
estimate, and that is what we want to obtain.

As there is no exact information available about the angle between
the spin and the magnetic axes for almost all of these isolated
neutron stars, at first we consider the low estimates of their
ages assuming constant inclination angle equal to maximal losses
position, $\tau_{\mathrm {cl}}$.
% assuming that the neutron stars remain in the
%region of active radio pulsars.

To get the lower age estimate for the case when sources are
already at the stage of extinct pulsars, $\tau_{\chi {\mathrm
{max}}}$, in the framework of the current losses model, we allow
the angle $\chi$ to evolve and find the maximum angle between the
magnetic and the spin axes at which the neutron star with the
presently observed spin period could still be located in the
region of active radio pulsars in the plane $P - \sin \chi$ (see
Fig.~7).\footnote{Of course, for PSR J1734-3333 this estimate
represent {\it the upper} age boundary.} The fact, that for
extinct pulsars in the scope of free particle escape model such an
estimate provides the lowest possible age is obvious.\footnote{We
do not consider the possibility that any of these NSs is on the
stage of orthogonal rotators, as all of them demonstrate
significant spin-down rates.}
 For the observed period an extinct pulsar has
$\chi
> \chi_{\mathrm{max}}$, so for the same initial period the one
which is now extinct has larger also initial inclination angle.
Then this means, that for all the same evolutionary parameters it
always had larger angle, and therefore, smaller $\dot P$. To reach
the same observed period from the same initial it takes longer for
the one which is now extinct. For the hindered particle escape
model we made numerical estimates, which demonstrate that even in
these case for all considered parameters, which have been chosen
according to parameters of the sources in Table~3, our value
$\tau_{\chi {\mathrm {max}}}$ still is a lower age estimate for
extinct radio pulsars.

%We want to emphasize that this age estimate does not
%include the possibility that the neutron star with the observed
%parameters has already passed to the region of extinct radio
%pulsars or to the stages of orthogonal rotators or supersonic
%propellers (at the stage of supersonic propeller the
%electrodynamic processes in the neutron star magnetosphere become
%insignificant, and the isolated neutron star spins down due to the
%interaction of the magnetosphere with the ambient matter).

\begin{figure}
\includegraphics[width=0.45\textwidth, keepaspectratio]{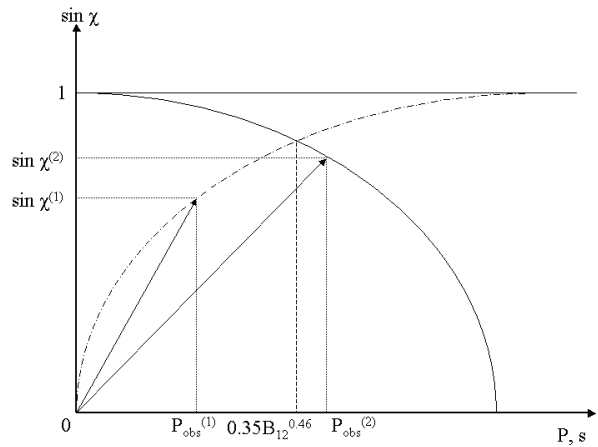}
 \caption{The maximum possible inclination angle $\chi_{\mathrm{max}}$ can be
defined as the angle at which the neutron star with the presently
observed period crosses the death line or the line of the orthogonal
rotators region. The death line (\ref{deathlinecl}) and the boundary
line of the region of orthogonal rotators (\ref{ortcl}) intersect at
the point with spin period $P_{\mathrm{int}} \approx 0.35
B_{12}^{0.46}$~s. If the presently observed spin period is less than
the period $P_{\mathrm{int}}$, the inclination angle
$\chi_{\mathrm{max}}$ is the angle at which the neutron star with
the presently observed period crosses the line of the orthogonal
rotators region. Otherwise, if the presently observed spin period is
longer than $P_{\mathrm{int}}$ then the maximum inclination angle
$\chi_{\mathrm{max}}$ is determined by the death line.}
\label{chimax}
\end{figure}

The age of the NS with $\chi= \chi_{\mathrm{max}}$ is calculated
according to Eq. (11), where $\chi_0=\chi_{\mathrm{max}}\,
P_0/P_{\mathrm{tr}}$. Here $P_{\mathrm{tr}}$ -- is the observed
period.

%We want to emphasize that the Table~3 is used solely to show how the
%evolution of an inclination angle affects the characteristic ages of
%neutron stars. The data listed in Table~3 do not represent the
%realistic age estimates since we use the specific assumptions about
%the inclination angles of these neutron stars, or at least they are
%not much more realistic than the one obtained under the standard
%assumption of $\sin \chi=1=$~const. Besides, an accurate procedure
%of age estimation should consider the allowance for a neutron star
%to pass to the region of extinct radio pulsars.

\begin{table*}
 \centering
 \begin{minipage}{\textwidth}
  \caption{Age estimates for different classes of isolated neutron
  stars in the scope of current losses model for the constant inclination
angle $\chi=0^{\circ}$ and for the inclination angle evolving to
the maximum possible value $\chi_{\mathrm{max}}$ . We assume that
a neutron star is located in the region of active radio pulsars on
the $P - \sin \chi$ diagram and the initial spin period is taken
to be $P_0=20$~ms.}
  \begin{tabular}{@{}lllllllllll@{}}
  \hline
Name & Class & P, s & B, G & $\tau_{\mathrm{cl}}$, Myr &
$\chi_{\mathrm{max}}$, deg & $\tau_{\mathrm{\chi
max}}$, Myr \\
\hline
 J1734-3333           & HB-PSR & 1.17 & $10^{14}$ & 0.05 & 89.2 & 0.1 \\
 1806-20              & SGR    & 7.5  & $10^{14}$ & 0.3 & 77.5 & 0.5 \\
 1E2259+586           & AXP    & 6.98 & $6 \times 10^{13}$ & 0.6  & 64.3 & 0.8 \\
 J0720.4-3125         & XDIN   & 8.39 & $10^{14}$ & 0.3  & 71.4 & 0.5 \\
  J1819-1458           & RRAT   & 4.26 & $5 \times 10^{13}$ & 0.5  & 83.7 & 0.9 \\
\hline
\end{tabular}
\end{minipage}
\end{table*}

All objects presented in Table 3 belongs to classes of highly
magnetized NSs with long spin periods. For them magneto-rotational
evolution goes quicker, so any evolution of $\chi$ on a spin-down
time scale should be very pronounced. Even without precise
information about inclination angles of highly magnetized NSs,
observations allow us to say that they cover a wide range: from
nearly orthogonal, to small $\chi$. From Tables 1 and 2 one can
see, that NSs with $B$ larger $\sim 10^{13}$ born with short
periods rapidly, in less than few tens of thousand years or even
faster, become aligned or orthogonal rotators without significant
spin-down, unless they were born with $\chi$ extremely close to
the critical value of maximal losses. As the latter does not seem
very probable for all of these sources (for such long periods and
strong fields the proximity to the critical values should be of
order of minutes), we can conclude that initial periods of highly
magnetized NSs should be close to the presently observed values.

Note that this conclusion is in correspondence with recent
investigation of properties of supernova remnants associated with
magnetars \citep{vk2006}. The authors do not find evidence that
these neutron stars were formed from rapidly rotating protoneutron
stars. Thus, it is likely that magnetars, as active radio pulsars,
can be born with comparatively long periods. Recent calculation
\citep{slss2006} support the conclusion that highly magnetized NSs
are born slowly rotating. Furthermore, \citet{fw2003,fw2006}
discussed that the neutron stars with strong magnetic fields tend
to be born as slow rotators, so initial spin period of a neutron
star may depend on the magnetic field strength. On the other hand,
for magnetars non-electromagnetic contributions to the evolution
of $\chi$ can be important.
%As it was note by \cite{macy1974},
%strong toroidal magnetic field result in counteralignment of spin
%and magnetic axes.

\subsection{Population synthesis}
\label{pops}

One of areas in which poor knowledge of the evolution of $\chi$
 can have strong impact is population synthesis of binary and isolated
 NSs. Below we mainly focus on the case of solitary objects, but in many aspects
 this discussion can be applied to binaries, too.

The inclination angle $\chi$ explicitly appears in the formulas for
spin-down rate and energy release in both models (magneto-dipole and
current losses). The death line should also depend on this
parameter, and in the case of the current losses model this is shown
explicitly. Finally, extreme values of this parameter can result in
low energy losses. Taken all together, it appears that modeling the
evolution of this parameter is at least as important as the
evolution of the magnetic field.

Most (if not all) of population synthesis models accept the
standard assumption about spin-down, which is Eq.(1) with $\sin
\chi =1=$const. The death line for radio pulsars is taken in the
form of Eq. (4). It is often believed that the uncertainty in the
initial values of $\chi$ can be hidden in uncertainties of $B$. In
some sense, this is true. But only when we deal with active radio
pulsars, and if the inclination angle does not significantly
evolve. On later evolutionary stages (propellers, accretors)
$\chi$ and $B$ ``decouples''. Such ``decoupling'' means that we
become interested in actual values of parameters, not in some
particular combination. In particular, to predict properties of
isolated accreting NSs it is very important to know if most of
them are expected to be aligned rotators (and so, non-pulsating
sources), or, oppositely, they tend to be orthogonal rotators
showing clear pulsations.

Also, the death line should depend on both $\chi$ and $B$, and at
least in the case of current losses the combination of these
parameters in the definition of the line is different from the
combination in the spin-down rate. The use of the standard
assumption, which, as demonstrated above, does not have solid
theoretical grounds, makes many results of population synthesis
not very trustable. So, if the aim is to produce a kind of a
global population synthesis, like the one presented in
\citep{p2000}, including post-PSR stages, then it is necessary to
follow the evolution of $\chi$ and $B$ separately. Unfortunately,
it has never been done, and there are serious difficulties in
fulfilling such a program, as none of theories gives such an
evolution in good correspondence with observations of radio
pulsars.

\section{Summary}
\label{summary}

% Мы рассмотрели магнито-вращательную эволюцию нейтронных звезд с
% учетом эволюции угла между магнитной осью и осью вращения, $\chi$.
% Рассмотрение проведено в рамках двух теорий: магнито-дипольных и
% токовых потерь. Результаты, полученные в рамках первой из них,
% не являются новыми, однако приводятся для сравнения и
% иллюстрации.
% Результаты расчетов сравниваются как с данными наблюдений, так и со
% стандартным предположением, широко использующимся при оценках
% возрастов нейтронных звезд и в популяционном синтезе.
%Последнее состоит в использование формулы магнито-дипольных потерь
%с постоянным углом $\chi$, равным 90 градусов.

 We revisited magneto-rotational evolution of NSs with account of
 the evolution of the angle $\chi$ between spin and magnetic axes.
 In our consideration we used two models of energy losses:
magneto-dipole and longitudinal currents. In general, the results
obtained in the framework of the former are not new, but are
presented for comparison and illustration. We compare output of
our calculations with observations, and with estimates according
to the standard assumption widely used for age estimates and in
population synthesis. In this assumption the spin-down is
calculated due to magneto-dipole model with constant
$\chi=90^{\circ}$.

%Теоретические расчеты показывают, что при начальных углах не
%слишком близких к критическим, соответствующим максимальным
%потерям (90 градусов в случае дипольных потерь, и 0 градусов в
%случае токовых), нейтронные звезды не переходят в область потухших
%пульсаров, пересекая линию смерти, а попадают в область малых
%потерь (соосные ротаторы в случае магнито-дипольных потерь, и
%ортогональные - в случае токовых). Причем, при малых начальных
%периодах и достаточно сильных (но еще не магнитарных) полях, эта
%эволюция происходит очень быстро, за десятки-сотни тысяч лет.
%Такая картина не соответствует данным наблюдений. Возможны три
%варианта разрешения противоречия.

 Our calculations demonstrate that for initial angles not too
 close to the critical ones ($90^{\circ}$ in the case of
 magneto-dipole losses, and $0^{\circ}$ -- for current losses)
 NSs do not become normal extinct pulsars crossing the death line,
 but appear in the region of low losses (aligned rotators in
 the case of magneto-dipole model, and orthogonal rotators in the
 case of current losses). At that, for small initial periods and
 strong (but still not magnetar) magnetic fields this evolution
 proceeds very rapidly, in tens or hundreds thousand years.
 This fact does not correspond with observations.
 We consider three possibilities to resolve this problem:

%1. Начальные периоды близки к наблюдаемым в настоящий момент

%2. В соответствии со стандартным предположением, начальные углы
%$\chi$ экстремально близки к критическим.

%3. Обе теории имеют существенные недостатки, исключающие их прямое
%использование для расчетов эволюции с учетом изменения угла
%$\chi$.

\begin{itemize}
\item an essential part of radio pulsar are born with comparatively long spin periods,
close to the observed ones, or
\item the initial angles between the magnetic and the spin axes are close
to critical values, or
\item both magnetodipole model and the model based on current losses require significant
modifications and cannot be directly applied for detailed
calculations of NS evolution.
\end{itemize}

%Для проверки первого предположения необходимо проведение
%детального популяционного синтеза. Однако представляется
%возможным, что результаты такого моделирования покажут
%неспособность данной гипотезы разрешить все противоречия,
%поскольку в расчетах должен появиться ненаблюдаемый избыток
%относительно старых пульсаров с большими (в случае токовых) или
%малыми (в случае магнито-дипольной модели) углами, кроме того,
%должен возрасти темп рождения пульсаров.

To test the first assumption it is necessary to perform a detailed
population synthesis modeling. However, it seems possible, that
the results of such computations would not support the hypothesis,
as we expect an unobserved overabundance of old pulsars with
angles corresponding to low energy losses. Also, the total birth
rate of pulsars should become higher, in contradiction with
independent estimates.

%Второе предположение, сводящееся к тому, что начальные углы $\chi$
%чрезвычайно (в пределах градуса) близки к одному из крайних
%положений представляется искусственным, и не имеет под собой
%физической основы. Оно явно должно быть модифицировано.

The second possibility that initial angles $\chi$ are very close
(within 1 degree) to one of the critical values seems too
artificial, and does not have a solid physical ground. Clearly, it
should be modified.

%Вероятнее всего, наше понимание эволюции нейтронных звезд на
%стадии радиопульсара на данный момент недостаточно, и обе теории
%не могут быть напрямую применены для детальных расчетов. Например,
%могут быть важны какие-то неэлекторо-магнитные эффекты, особенно
%на ранних стадиях торможения, связанные с наличием дисков вокруг
%нейтронных звезд или с механическими эффектами, связанными с
%несферичностью или внутренними процессами в компактных объектах.
%Эти эффекты и их вклады могут быть различны для разных типов
%нейтронных звезд.

Most probably, our understanding of NS evolution on the radio pulsar
stage in not satisfactory, now, and so both theories cannot be
directly applied for detailed computations. Some non-electromagnetic
effects can be important, especially during early stages of
spin-down. They can be related to the presence of debris discs
around NSs, or with mechanical effects related to non-sphericity or
to processes in NS interiors. These effects can be different for
different types of NSs.

%С точки зрения проверки предсказаний теории и выявления начальных
%параметров наиболее перспективными представляются
%сильнозамагниченные нейтронные звезды. Их эволюция протекает
%наиболее быстро, для них есть оценки углов $\chi$ и независимые
%оценки возраста. Наши оценки (включая реконструкцию треков)
%показывают, что, если такие источники как SGRs, AXPs, RRATs, M7
%находятся за линией смерти, то их начальные периоды вращения
%должны быть близки к наблюдаемым, в том случае, если верна одна из
%рассмотренных теоретических моделей, и вообще, если угол
%существенно эволюционирует на временном масштабе изменения
%периода.

Highly magnetized NSs seem to be most favorable sources to test
predictions of theories of magneto-rotational evolution, and to
probe initial parameters. The evolution of such stars is the most
rapid, for them there are estimates of angles $\chi$ and
independent ages estimates. Our calculations (including track
reconstructions) show that if such sources as SGRs, AXPs, RRATs,
and the Magnificent seven are beyong the death line, then their
initial spin periods should be close to the observed ones. This
conclusion is valid if one of the discussed theories is working,
or, in general, if the inclination angles significantly evolve on
the spin-down time scale.

%Незнание деталей эволюции, связанных с поведением угла $\chi$,
%наиболее существенно для оценок возрастов отдельных нейтронных
%звезд и для моделей популяционного синтеза. Учитывая значительную
%роль угла $\chi$ в обеих моделях энергопотерь нейтронных звезд
%(равную роли величины магнитного поля), можно сказать, что
%игнорирование эволюции этого параметра является "скелетом в шкафу"
%моделей популяционного синтеза.
% Учет этой эволюции крайне важен при
%популяционном синтезе радиопульсаров и существенно влияет на
%расчеты более поздних стадий эволюции (пропеллеры, аккреторы).

Poor understanding of the details of evolution related to changes
of the inclination angle is mostly important for age estimates and
for population synthesis studies. Taking in account important role
of $\chi$ in both models of energy losses (at least, equal to the
role of the magnetic field strength) we can say, that neglect of
its evolution is a ``skeleton in a cupboard'' of population
synthesis models. The account of the evolution of $\chi$ can be
crucial for radio pulsar studies, and significantly impact
calculations of the later stages (propellers, accretors, etc.).

%Резюмируя, можно сказать, что наше понимание магнито-вращательной
%эволюции нейтронных звезд недостаточно для построения детальных
%популяционных моделей и точных оценок возраста. Использование
%стандартного предположения о постоянстве угла $\chi$ противоречит
%имеющимся на сегодняшний день теориям энергопотерь пульсаров, и
%нуждается в совершенствовании или обосновании. По всей видимости,
%детальное численное моделирование
%\citealt{timokhin2006,spitkovsky2006} будет способно дать ответ на
%вопрос о деталях магнито-вращательной эволюции нейтронных звезд.

Summarizing, we can say that our understanding of NS
magneto-rotational evolution is not enough to make detailed
population models and to obtain exact ages. The use of the standard
assumption of constant $\chi$ contradicts present day models of
radio pulsar energy losses and require serious modifications or
basis. Most probably, detailed numerical modeling
\citep{timokhin2006,spitkovsky2006} can provide the correct answer
about details of magneto-rotational evolution, in the future, but,
probably, additional inclusion of non-electromagnetic processes can
be necessary.

\section*{Acknowledgments}

We thank the unknown referee for useful comments and criticism
which helped to improve the paper.
 SBP is grateful to Roberto
Turolla for useful discussions. In this research we used the data
from the ATNF Pulsar Catalog provided by ATNF Pulsar Group, and
AXP/SGR Online Catalog provided by McGill Pulsar Group. SAE and
VSB acknowledge support from the grant RFBR 05-02-17700. The work
of SBP was supported by the RFBR grant 04-02-16720 and by the
''Dynasty'' Foundation (Russia).
 SBP is the CARIPLO Foundation Fellow.

%\renewcommand{\thefigure}{\arabic{figure}}
%\setcounter{figure}{2}

%\renewcommand{\thefigure}{\arabic{figure}\alph{subfigure}}
%\setcounter{subfigure}{1}

%\addtocounter{figure}{-1}
%\addtocounter{subfigure}{1}

\end{document}